\documentclass[%
 reprint,
  superscriptaddress,
 amsmath,amssymb,amsfonts,
 aps,
pra,
 floatfix,
 longbibliography
]{revtex4-2}

\usepackage[utf8]{inputenc}
\usepackage[pdftex]{graphicx} \graphicspath{{}}
\usepackage{float} \usepackage{color}
\usepackage[pdftex,colorlinks=true]{hyperref}
\usepackage{subfigure}
\hypersetup{
  colorlinks=true,
  linkcolor=blue,
  citecolor = blue,
  urlcolor=blue,
}
\usepackage[utf8]{inputenc}
\usepackage{amsmath}

\newcommand{\non}{\nonumber}
\newcommand{\beq}{\begin{equation}}
\newcommand{\eeq}{\end{equation}}
\newcommand{\bea}{\begin{eqnarray}}
\newcommand{\eea}{\end{eqnarray}}
\usepackage{mathtools}

\DeclarePairedDelimiterX\braket[2]{\langle}{\rangle}{#1 \delimsize\vert #2}
\DeclarePairedDelimiterX\expval[3]{\langle}{\rangle}{#1 \delimsize\vert #2  \delimsize\vert #3}
\DeclarePairedDelimiterX\proj[2]{\delimsize\vert#1\rangle}{\langle#2\delimsize\vert}{ }
\usepackage{siunitx}

\usepackage{tikz}

\begin{document}

\title{Interacting Bose gases in twisted-bilayer optical lattices}
\author{Ganesh C. Paul}
\affiliation{Institut f\"{u}r Mathematische Physik, Technische Universit\"{a}t Braunschweig, D-38106 Braunschweig, Germany}
\author{Patrik Recher}
\affiliation{Institut f\"{u}r Mathematische Physik, Technische Universit\"{a}t Braunschweig, D-38106 Braunschweig, Germany}
\affiliation{Laboratory for Emerging Nanometrology Braunschweig, D-38106 Braunschweig, Germany}
\author{Luis Santos}
\affiliation{Institut f\"{u}r Theoretische Physik, Leibniz Universit\"{a}t, 30167 Hannover, Germany}
\date{\today}

\begin{abstract}
Recent experiments have realized ultra-cold gases in twisted-bilayer optical lattices. 
We show that interacting bosons in these lattices present a highly non-trivial ground-state 
physics resulting from the interplay between inter- and intra-layer hopping and interactions.
This physics is crucially determined by site clusterization, which we properly take into account 
by developing a specifically-tailored cluster Gutzwiller approach. Clusterization results 
in a large variety of different Mott-like phases characterized by typically different 
occupations of the clusters, and in the appearance of pockets of sites in between which particles can 
freely move, but which remain disconnected from each other. 
This peculiar phase, which resembles the well-known Bose glass phase, may 
occur even for commensurate twist angles and is further enhanced when the twisting is incommensurate. Moreover, in the 
incommensurate case, the formation of mobility islands may occur even without inter-layer hopping solely due to inter-layer interactions.
\end{abstract}

\maketitle
\section{Introduction}

Recent years have witnessed major breakthroughs 
in solid state studies of twisted-bilayer materials and moiré physics~\cite{bistritzer2011moire, andrei2020graphene,tarnopolsky2019origin}, including the observation of anomalous 
supercondutivity~\cite{cao2018a,yankowitz2019,oh2021,yankowitz2019tuning} and insulating behavior~\cite{cao2018correlated,codecido2019correlated,nuckolls2020strongly,cao2020tunable}. Whereas at magic commensurate twisting angles 
a periodic superlattice~(moir\'e) modulation develops~\cite{bistritzer2011moire}, 
at incommensurate twistings the system presents quasi-disorder, resembling quasicrystals~\cite{shechtman1984metallic}.

Recently, ultracold atoms in optical potentials have been revealed as a suitable platform for the quantum simulation of 
twisted-bilayer physics
~\cite{gonzalez2019cold,
salamon2020simulating,luo2021spin,lee2022emulating,madronero2023dynamic,paul2023particle, wan2024fractal}. Twisted-bilayer square lattices using two coupled internal states as effective bilayers in a synthetic dimension~\cite{gonzalez2019cold} have been recently realized by employing atomic Bose–Einstein condensates loaded into spin-dependent optical lattices ~\cite{meng2023atomic}. Optical lattice experiments, which are characterized by an exquisite experimental control of 
all relevant parameters~(most relevantly the interparticle interactions, hopping amplitudes, and the lattice geometry) open exciting perspectives for the study of the properties of interacting particles in twisted potentials. 

Short-range interacting bosons in regular periodic lattices present as a function of the chemical potential and hopping rate, either a superfluid or a Mott insulator phase in the ground-state~\cite{greiner2002quantum}. The presence of disorder or quasi-disorder 
results in the appearance of the Bose glass~(BG)~\cite{fisher1989boson, lugan2007ultracold, fallani2007ultracold}, an insulator which, in contrast to the Mott phase, presents a finite compressibility. Recently, Bose glasses have been discussed in the context of quasi-crystalline structures
~\cite{johnstone2021mean,ciardi2023quasicrystalline,yu2024observing}, 
including single-layer twisted potentials, where weak superfluidity was reported for commensurate angles~\cite{johnstone2024weak}.
Very recently, it was shown that bosons in twisted-bilayer optical lattices may present in the absence of interlayer hopping 
a BG for commensurate angles, due to the presence of inter-layer interactions~\cite{zeng2024dynamical}.



\begin{figure}[t!]
\centering
\includegraphics[width=\columnwidth]{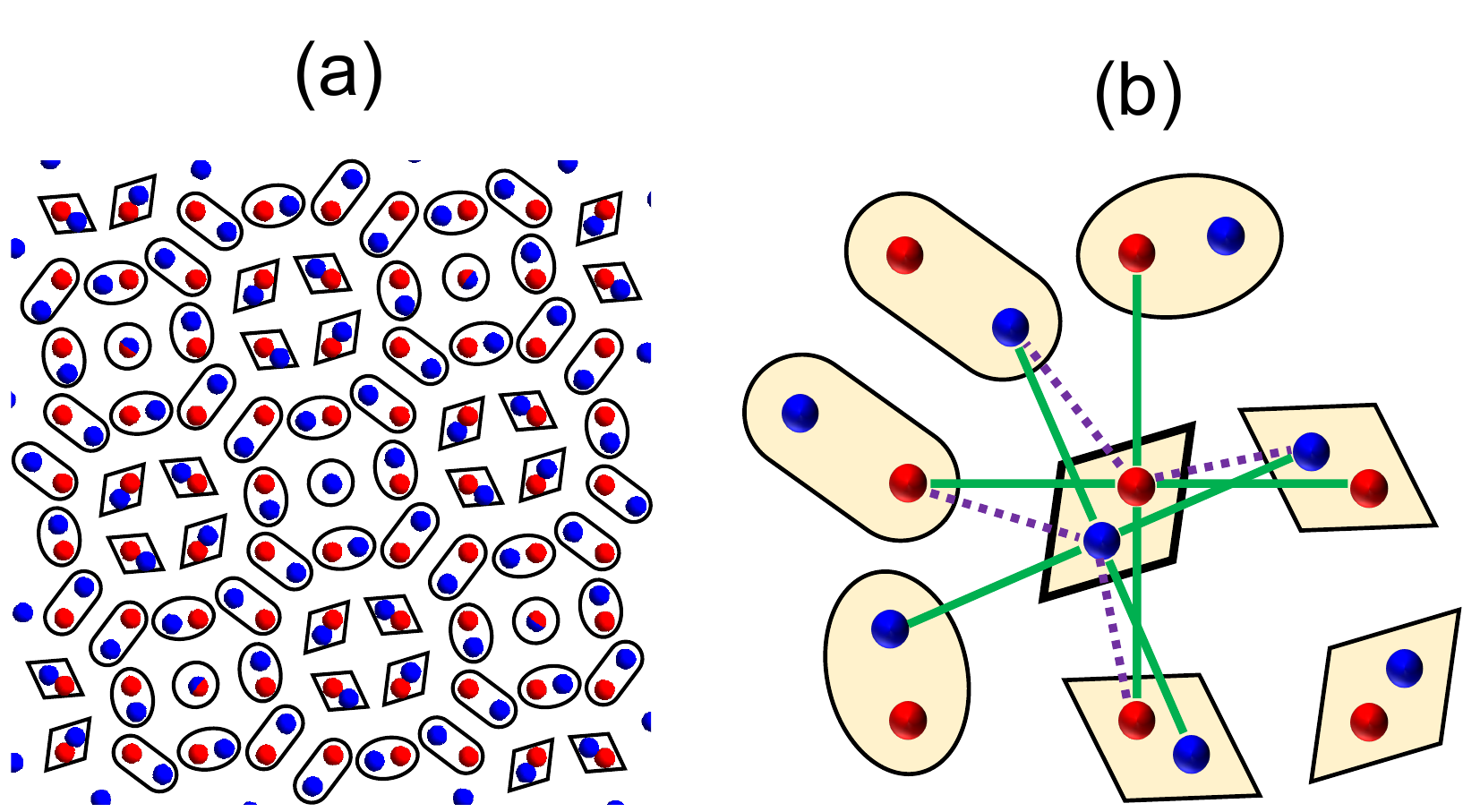}
\caption{(a) Twisted-bilayer lattice for 
a moir\'e angle $\theta=\theta(3,2)$. Sites of layer 1~(2) are indicated by filled red~(blue) circles. We indicate with different forms the different cluster types: cluster $1$~(circle), $2$~(rhombus), $3$~(oval) and $4$~(rounded rectangle). These clusters are characterized by different inter-layer couplings, see text. For this lattice there are no unclustered sites. (b) Hopping between neighboring clusters results from intra-layer~(inter-layer) tunneling indicated with solid-green~(dashed-purple) lines.}
\label{fig:1}
\end{figure}


In this paper, we investigate the ground-state properties of interacting ultra-cold bosons in 
a twisted-bilayer optical lattice. Due to inter-layer tunneling, the lattice splits into clusters of two sites, each one belonging to a 
different layer, which are connected by a significant inter-layer hopping. We develop a specific Gutzwiller Ansatz that 
takes properly into account clusterization. Equipped with this method, we show that the interplay between 
interactions and the twiste-induced lattice clusterization results in a rich ground-state physics, for both 
commensurate and incommensurate twistings. Mott-like phases are 
characterized by a non-trivial dependence of the filling of the different clusters with the chemical potential, resulting typically 
in a density wave modulation. Bose-glass phases may occur even for commensurate angles due to the formation of "quantum wheels" similar to those observed in quasi-crystals in Ref.~\cite{ciardi2023quasicrystalline}. For incommensurate twistings, the 
Bose glass is characterized by the formation of non-trivial isolated domains of sites, in between which particles may freely move. 
For sufficiently large hopping amplitudes, the eventual percolation of these islands leads to a superfluid phase~\cite{niederle2013superfluid,johnstone2021mean}. 

The paper is organized as follows. In Sec.~\ref{sec:Model} we introduce the model, the cluster Gutzwiller approach employed throughout the paper, as well as the percolation analysis. Section~\ref{sec:Commensurate} is devoted to the case of 
commensurate twistings and the formation of mobility islands. Section~\ref{sec:Incommensurate} deals with incommensurate 
twisting angles, discussing separately the case with and without inter-layer hopping. Finally, in Sec.~\ref{sec:Conclusions} we summarize our conclusions.


\section{Model and Methods}
\label{sec:Model}

\subsection{Hamiltonian}

In the following, we consider contact-interacting bosons in two layers of square optical lattices, where one layer is twisted by an angle $\theta$ with respect to the other. We focus in particular on the case in which this arrangement is engineered in the way proposed in Ref.~\cite{gonzalez2019cold}, and recently realized experimentally in Ref.~\cite{meng2023atomic}, in which 
two internal levels of the atoms act as two effective layers in a synthetic dimension, whereas 
a state-dependent potential results in the twisted-bilayer geometry. For a square lattice geometry, a periodic moir\'e pattern is obtained for twist angles $\theta=\theta(m,n)\equiv\arccos(\frac{2mn}{m^2+n^2})$ ($n,m\in\mathbb{Z}$).  The Hamiltonian of the 
system acquires the form:
\bea
H&=&-t \sum_{\alpha}\sum_{\langle j,j'\rangle} a_{\alpha, j}^\dagger a_{\alpha, j'}
-\sum_{\alpha\neq\bar{\alpha}}\sum_{j,j'} t_\perp (j,j')a_{\alpha, j}^\dagger a_{\bar{\alpha}, j'}\non\\
&+&\frac{U}{2}\sum_{\alpha}\sum_ j\hat n_{\alpha,j} 
(\hat n_{\alpha,j} - 1) - \mu \sum_{\alpha}\sum_j  \hat n_{\alpha,j}\non\\
&+& \sum_{\alpha}\sum_{jj'} V_\perp(j,j') \hat n_{\alpha,j} \hat n_{\bar{\alpha},j'}
\label{Ham}
\eea
where $a_{\alpha, j}$~($a^\dagger_{\alpha, j}$) is the annihilation~(creation) operator at site $j$ of layer $\alpha=\{1,2\}$, $\hat n_{\alpha,j}=\hat a_{\alpha,j}^\dag \hat a_{\alpha,j}$, $U$ characterizes the on-site repulsive interactions, assumed to be the same in both layers, and $\mu$ is the chemical potential. Note that whereas
intra-layer hopping to nearest neighbors is given by a fixed hopping rate $t$, tunneling between a site $j$ in one layer and site $j'$ in the opposite one is characterized by the variable rate $t_\perp (j,j')$. The repulsive inter-layer interactions are given by the couplings $V_\perp(j,j^{'})$. In the considered scenario using synthetic dimensions, intra-layer hopping 
results from either a microwave or two-photon optical Raman coupling between the two internal levels, whereas inter-layer interactions result from contact-like interactions between the two internal components. For a sufficiently deep optical lattice, the on-site Wannier functions (in each layer) can be approximated as a Gaussian of width $l_0=\frac{a}{\pi s^{1/4}}$, where $a$ is the lattice spacing characterizing the square lattice in both layers, and $s$ is the lattice depth in recoil units. As a result, the distance-dependent inter-layer hopping and interaction acquire, respectively, the form~\cite{gonzalez2019cold}: $t_\perp(j,j^{'})=t_\perp e^{-|R_{1j}-R_{2j^{'}}|^2/{4l_0^2}}$, and $V_\perp(j,j^{'})=V_\perp e^{-|R_{1j}-R_{2j^{'}}|^2/{2l_0^2}}$.  We have chosen $l_0=0.15a$~($s\simeq 20$) throughout the paper. 
Note that $t_\perp$ can be controlled by the strength and detuning of the coupling between internal levels, whereas
$V_\perp$ depends on the contact-like inter-component interactions.

\subsection{Site clusterization}

 Due to the twisting between the two lattices, and the Gaussian dependence of the inter-layer hopping, sites at different positions present a markedly different coupling to sites in the opposite layer. In particular, pairs of neighboring sites $i$ and $j$ may present a significant inter-layer hopping $t_\perp(i,j)$, much stronger than the inter-layer coupling to any other site. Moreover, for sufficiently low $t$, $t_\perp(i,j)$ may overwhelm as well the intra-layer tunneling. When this occurs, in first approximation, those two sites could be then considered as a cluster decoupled from the rest of the lattice, forming an effective two-well potential. As a result, a single particle in those two sites would occupy in the ground-state the symmetric superposition $(|i\rangle + |j\rangle)/\sqrt{2}$ with energy $-t_\perp(i,j)$. Although this decoupling is generally not exact, and inter-cluster hopping plays a relevant role (as depicted in Fig.~\ref{fig:1}~(b)), site clusterization turns out to be crucial to understand the ground-state physics of the interacting Bose gas in the twisted bilayer lattice.  

 We define a cluster $c$ as formed by two sites $c_1$ and $c_2$, in layers $1$ and $2$, respectively, such that $t_\perp(c_1,c_2)>t_{\perp,cr}$. In the following, we set $t_{\perp,cr}=0.03$~$(0.06)t_\perp$ for commensurate~(incommensurate) twist angles. We have checked that lower values of $t_{\perp,cr}$ do not appreciable change the results discussed below. Note that due to the Gaussian dependence of the inter-layer tunneling, for all values of $t_\perp$ discussed in this paper, clusters are always limited to two sites. In addition to the clustered~(cl) states, there may be also sites without a significant coupling to any neighboring site in the opposite ladder. They do not belong to a cluster, and are referred below as non-clustered~(ncl) sites. For the case of a given moir\'e angle, the clusters can be classified in well-defined types, characterized by a different inter-layer coupling. The case of  $\theta=\theta(3,2)$ is shown in Fig.~\ref{fig:1}~(a),  where we indicate the different cluster types. As discussed below, coupling between neighboring clusters via intra- and inter-layer tunneling, illustrated in Fig.~\ref{fig:1}~(b), plays a key role in the eventual on-set of superfluidity in the lattice.
 
Motivated by the previous discussion, we introduce the
Gutzwiller-like ansatz $|\Psi\rangle =  |\Psi_{cl}\rangle \otimes |\Psi_{ncl}\rangle$. Clustered sites are described by $|\Psi_{cl}\rangle = \bigotimes_{c \in cl} |\rho_c^{cl}\rangle$, with 
$|\rho_c^{cl}\rangle = 
\sum_{n_1, n_2} g_c(n_1,n_2) |n_1\rangle_{c_1} |n_2\rangle_{c_2}$, where $g_c(n_1,n_2)$ determines the amplitude of probability of finding $n_1$~($n_2$) particles in site $c_1$~($c_2$) of cluster $c$. Non-clustered sites are characterized by $|\Psi_{ncl}\rangle = \bigotimes_{\alpha; j \in ncl} |\rho_{\alpha, j}\rangle$, with 
$|\rho_{\alpha j}\rangle = \sum_n f_{\alpha j}(n) |n\rangle_{\alpha j}$, where $f_{\alpha,j}(n)$ is the amplitude of probability of finding $n$ particles in the non-clustered site $j$ of layer $\alpha$. The coefficients fulfill the normalization conditions: $\sum_n |f_{\alpha j}(n)|^2 =\sum_{n_1,n_2} |g_c(n_1,n_2)|^2 =1$. 

An uncoupled non-clustered site $(\alpha,j)$ is characterized by an energy:
\begin{equation}
\tilde{E}_{(\alpha,j)}= \sum_{n } \epsilon(n) \left |f_{\alpha, j}(n)\right |^2, 
\end{equation}
with $\epsilon(n)=\frac{U}{2}n(n-1) -\mu n$, whereas
the energy of an uncoupled cluster c is:
\begin{eqnarray}
&& \!\!\!\!E_{c} =  \sum_{n_1, n_2} 
\xi_c(n_1,n_2)\left |g_c\left (n_1,n_2 \right ) \right |^2  \\
&-& t_{\perp,c}\!\! \sum_{n_1, n_2} \!\!\!\left (\!\sqrt{n_1(n_2\!+\!1)}  \,  g_c^{*}(n_1,n_2) g_c(n_1\!-\!1,n_2\!+\!1)\!+\!\mathrm{c.c.} \!\right )\!,\non
\end{eqnarray}
with $t_{\perp,c}$ and $V_{\perp, c}$ characterizing the inter-layer couplings in the cluster, and  $\xi_c(n_1,n_2)=
\sum_{\sigma={1,2}} 
\epsilon(n_\sigma)+V_{\perp,c} n_1 n_2$. 

Neglecting interactions between different clusters or unclustered sites~(which is extremely small under the conditions discussed below), coupling between clusters 
is only given by hopping. Intra-layer hopping between neighboring clusters $c$ and $c'$ result in the energy $T_{c,c'}^{\mathrm{intra}}=-t\sum_\alpha (b_{\alpha,c_\alpha}^* b_{\alpha,c'_\alpha} +\mathrm{c.c.})$, if $c_\alpha$ and $c'_\alpha$ are nearest neighbors in layer $\alpha$, whereas inter-layer 
tunneling results in the contribution: $T_{c,c'}^{\mathrm{inter}}=-t_\perp(c_\alpha,c'_{\bar{\alpha}}) (b_{\alpha,c_\alpha}^*b_{\bar{\alpha},c'_{\bar{\alpha}}} + \mathrm{c.c})$. Similarly, the hopping between two neighboring ncl sites contributes with $T_{(\alpha,i),(\alpha,j)}^{\mathrm{intra}} = -t b_{\alpha,i}^*b_{\alpha,j} + \mathrm{c.c}$, and   $T_{(1,i),(2,j)}^{\mathrm{inter}} = -t_\perp(i,j) b_{1,i}^*b_{2,j} + \mathrm{c.c}$. Finally, 
hops between clustered and non-clustered sites result, respectively, in 
couplings $T_{(\alpha,c),(\alpha,j)}^{\mathrm{intra}}=-t b_{\alpha,c_\alpha}^*b_{\alpha,j}+\mathrm{c.c.}$, where 
$c_\alpha$ and $j$ are nearest neightbors, and 
$T_{(\alpha,c),(\bar{\alpha},j)}^{\mathrm{inter}}=-t_\perp(c_\alpha,j) (b_{\alpha,c_\alpha}^*b_{\bar{\alpha},j} + \mathrm{c.c.})$. In the previous expressions we have introduced the mean-fields:
$b_{\alpha,c\in \mathrm{cl}} \equiv \langle \hat a_{\alpha,c} \rangle$ and $b_{(\alpha, j)\in \mathrm{ncl}} \equiv \langle \hat a_{\alpha,j} \rangle$, which are determined from the Gutzwiller coefficients:
\bea
b_{1,c} &=& \sum_{n_1, n_2} \sqrt{n_1+1} \,\, g_c^{*}(n_1,n_2)g_c(n_1+1,n_2), \\
b_{2, c} &=&  \sum_{n_1, n_2}\sqrt{n_2+1} \,\, g_c^{*}(n_1,n_2)g_c(n_1,n_2+1), \\
b_{\alpha, j} &=& \sum_n \sqrt{n+1} \,\, f_{\alpha, j}^{*}(n) f_{\alpha, j}(n+1).
\eea
Adding up all the energy terms, we obtain the overall energy $E$, which we minimize obtaining the coefficients 
$g_c(n_1,n_2)$ and $f_{\alpha,j}(n)$.

\subsection{Percolation analysis}

In Bose–Hubbard models in uniform square lattices, the superfluid~(SF) phase is characterized by a non-zero mean-field 
$\langle \hat a_j \rangle$ in all sites $j$, whereas a Mott-like insulator~(MI) present a zero mean-field and 
an integer $\langle \hat n_j \rangle$.
In the twisted-bilayer scenario, 
sites belonging to different clusters (or sites in non-cluster states) typically present different fillings and different values of the mean fields $b_{\alpha,s}$. As discussed below, this leads to a rich landscape of possible phases.

Due to the inherent inhomogeneity of twisted-bilayer lattices, some sites may present a vanishing mean-field, whereas others may have a non-vanishing one. In our analysis, we fix the cut-off $|b_{\alpha,s}|<0.05$ to determine whether a site has a vanishing mean-field. 
We associate at this point the parameter $S=1$~($0$) to each site with non-vanishing~(vanishing) mean-field.

We define a Mott-like phase as having $S = 0$ for all sites, and hence clusters with an integer filling. Note however that different clusters can have different integer fillings, hence resulting rather in a density wave. When at least some sites present $S=1$, superfluidity is only established if $S=1$ sites percolate~\cite{niederle2013superfluid, hettiarachchilage2018local, johnstone2021mean}. 
We determine the network of sites formed by connecting all the nearest-neighboring~(both intra- and inter-layer) sites with $S=1$. We consider two $S=1$ sites as connected if their coupling is larger than $10^{-3}t_\perp$. A slight change of this value does not affect any of our results. We then employ a percolation analysis, based on the Hoshen-Kopelman algorithm~\cite{hoshen1976}, to determine whether there is at least one chain of $S=1$ sites that percolates from one side to the other of our system. 
Note that although percolation may not occur, there may be a significant fraction of neighboring sites with 
$S=1$, which will then form an isolated superfluid region. This phase, as discussed below, may occur even in the case of commensurate moiré angles.

Motivated by the previous discussion, and following Refs.~\cite{niederle2013superfluid,johnstone2021mean},  we introduce two further parameters: the fraction of $S=1$ sites $F\equiv N_\phi/N$, and the percolation probability $P\equiv N_{\mathrm{span}}/N$, where $N_{span}$, $N_\phi$ and $N$ denote, respectively, the number of $S=1$ sites in a percolating chain, the total number of $S=1$ sites, and total number of sites in the two layers. Mott-like phases are characterized by $F=P=0$, whereas 
for the SF phase $F,P>0$. Non-percolated phases with $\langle S \rangle >0$ present $P=0$ but $F>0$. In the following, and due to their resemblance with the Bose glass phase known in disorder Bose-Hubbard models~\cite{fisher1989boson}, we denote these phases as BG phases.


\section{Commensurate twist angle}
\label{sec:Commensurate}

We consider first the case of a commensurate twist angle $\theta=\theta(m,n)$. We may then simplify our analysis by using periodic boundary conditions~(PBC), focusing on a single unit cell of the effective moiré lattice.
We illustrate the possible ground-state phases of the interacting Bose gas with the case of $\theta=\theta(3,2)$, since it presents already a complex landscape of clusters, but the different phases are still intuitively easy to visualize. The lattice presents four different cluster types, $c=1\dots 4$, see Fig.~\ref{fig:1}. Clusters $1$~(circles) are formed by two sites on top of each other, and hence $t_{\perp,c=1}/t_\perp=1$. The rest of the clusters 
present a lower inter-layer coupling: 
cluster $2$~(rhombuses) present 
$t_{\perp,2}/t_\perp\simeq 0.425$, 
cluster $3$~(ovals) $t_{\perp,3}/t_\perp\simeq 0.181$, and cluster $4$~(rounded rectangles)
$t_{\perp,4}/t_\perp\simeq 0.033$.



\begin{figure}[t!]
\centering
\includegraphics[width=0.8\columnwidth]{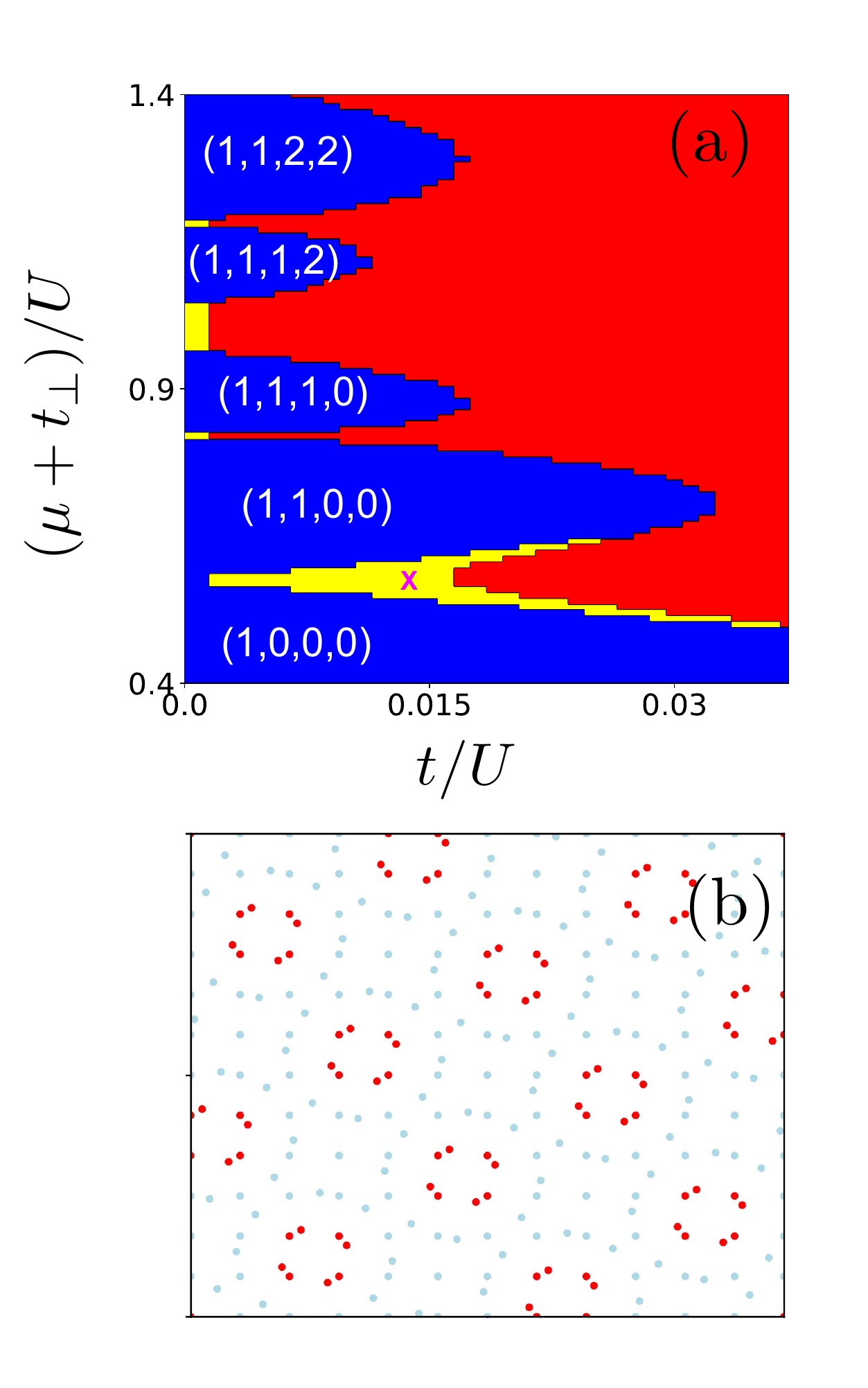}
\caption{(a) Phase diagram for $\theta=\theta(3,2)$, $l_0=0.15a$, $t_\perp=U$, and $V_\perp=4U$. Insulating Mott-like phases are indicated in blue, SF phases in red, and BG phases in yellow.
The Mott-like phases are labelled 
in the form $(N_1, N_2,N_3,N_4)$, where $N_c$ is the population of cluster $c$~(see Fig.~\ref{fig:1}). 
(b) Distribution of sites with $S=1$~(red) and $S=0$~(light blue) characteristic of the BG phase in between $(1,0,0,0)$ and $(1,1,0,0)$~(for the case indicated with a cross in Fig.~(a)).}
\label{fig:2}
\end{figure}


Figure ~\ref{fig:2}~(a) shows the phase diagram for $t_\perp=U$, and  $V_\perp=4U$~(other choices lead to qualitatively similar phase diagrams).
The Mott-like phases~(indicated in blue) are characterized by an integer, generally different, population $N_c$ of the different cluster types. We 
denote these phases in the form
$(N_1, N_2, N_3, N_4)$. 

At low chemical potential, the lattice is first filled 
with one particle in each cluster $1$, due to the strong inter-layer hopping rate in these clusters. Increasing the chemical potential results in the filling of other clusters, starting with cluster $2$. Note however that the transition between the $(1,0,0,0)$ and $(1,1,0,0)$ phases occurs at low $t/U$ via an intermediate phase, illustrated in Fig.~\ref{fig:2}~(b), in which the dc sites have unit filling, but clusters $2$ form superfluid ring islands, which resemble the quantum wheels recently reported in the context of quasi-crystal potentials~\cite{ciardi2023quasicrystalline}. These islands remain however isolated, without percolating through the lattice. As a result the system is in a BG phase~(indicated in yellow), with $P=0$ but $F>0$. At larger $t/U$ the coupling with clusters $3$ eventually results in a percolating network of clusters $2$ and $3$, and hence the system enters the SF phase~(red). Larger values of the chemical potential result in a non-trivial series of Mott-like phases, in which the filling is not necessarily sequential, e.g. $(1,1,1,0)$ is followed by $(1,1,1,2)$, with an appreciable gap 
in between. Concerning that gap, note that even if $t=0$, mobility can be established exclusively via inter-layer hops, explaining the absence of Mott phases at $t/U=0$ when $(\mu+t_\perp)/U\simeq 1$.



\begin{figure}[t!]
\centering
\includegraphics[width=0.5\columnwidth]{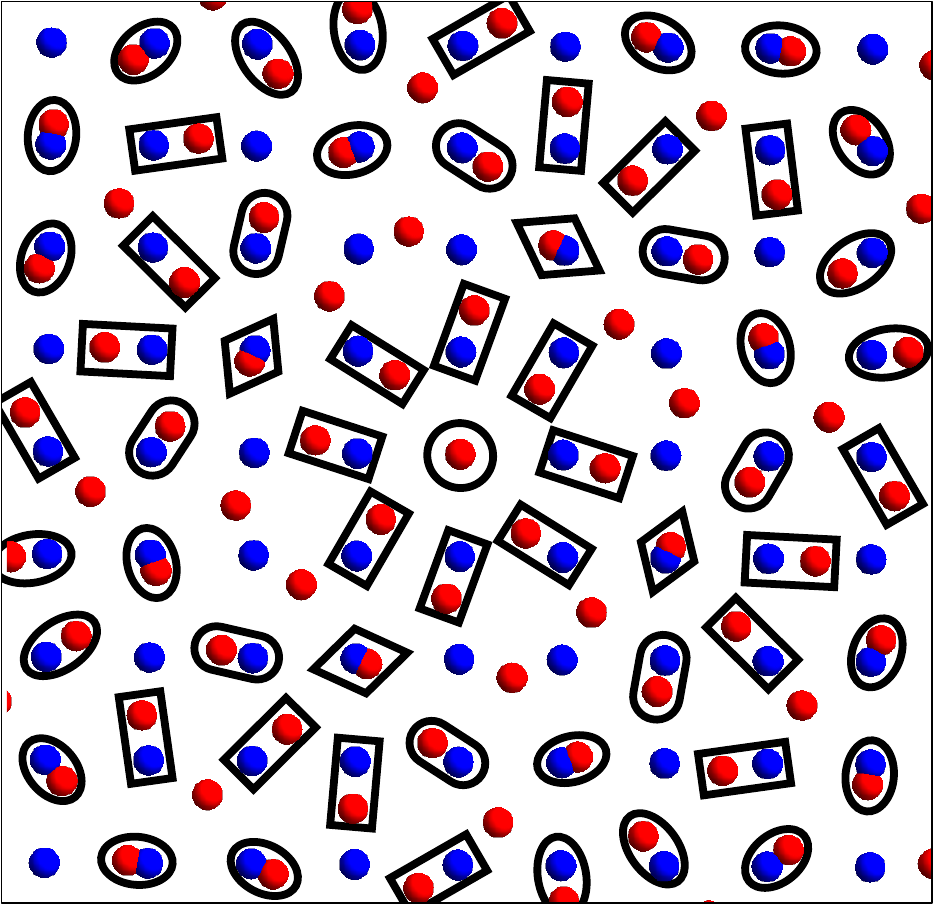}
\caption{
Spatial distribution of sites in layer 1~(blue) and layer 2~(red) for a lattice with an angle 
$\theta(2,1)+3^\circ$. For simplicity, in addition to the 
two central overlapping sites~(circle), we classify the rest of the clusters in four families: 
$1$~(rhombuses), $2$~(ovals), $3$~(rounded rectangles), and $4$~(rectangles). Note also the presence of non-clustered sites.}
\label{fig:3}
\end{figure}

\begin{figure*}[t!]
\centering
\includegraphics[width=2.0\columnwidth]{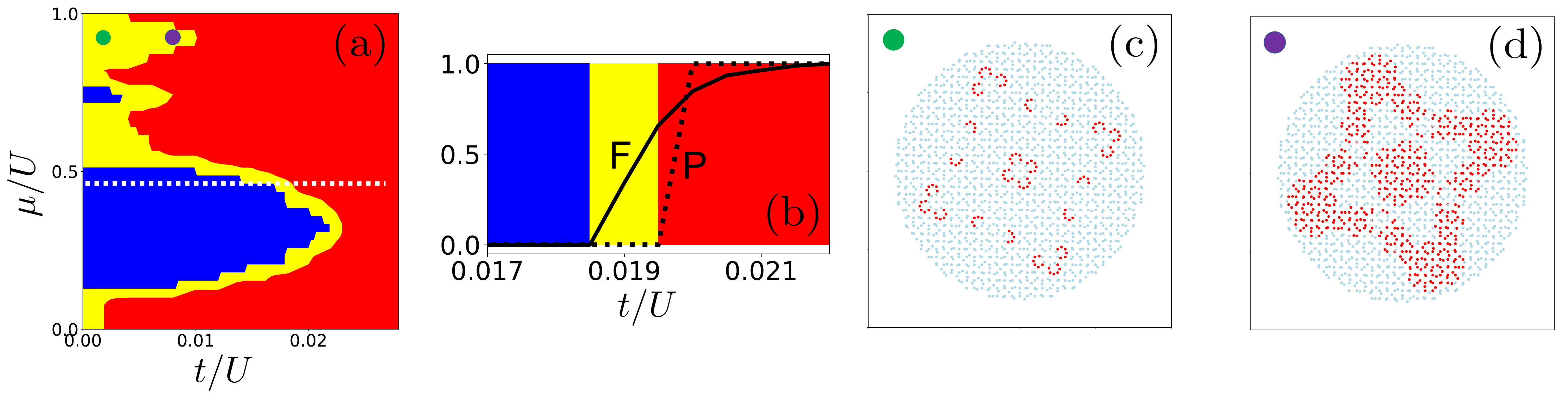}
\caption{(a) Phase diagram for $\theta = \theta(2,1)+3^\circ$, $t_\perp/U=1$, and $V_\perp=0$. 
Insulating Mott-like phases are indicated in blue, BG phases in yellow, and superfluid phases in red.
(b) the $F$~(solid line) and $P$~(dashed line) parameters~(see text) as a function of $t/U$ for $\mu/U=0.45$. These parameters are employed to characterize the phases.
(c)-(d) distribution of sites with $S=1$~(red) and $S=0$~(light blue) for the cases indicated 
in (a). We choose $40\times40$ sites in each layer.
}
\label{fig:4}
\end{figure*}


\section{Incommensurate twist angle}
\label{sec:Incommensurate}

We consider next the case of incommensurate twist angles. Since the lattice lacks periodicity, we cannot employ periodic boundary conditions and must hence work with finite-size systems. Inter-layer clusters of two neighboring sites may still be formed. Sites which experience an inter-layer coupling $t_{\perp,c}>0.06 t_\perp$ are considered as belonging to the same cluster. This cut-off value prevents in the cases discussed below the existence of clusters of more than two sites, without compromising the generality of the results. Sites which experience an interlayer coupling below that cut-off are considered as non-clustered sites. Our numerical simulation is performed 
on a $40\times 40$ twisted bilayer optical lattice.

In the following, we illustrate the possible ground-state phases for the case of incommensurate twist angles, using the relatively simple case of 
$\theta=\theta(2,1)+3^\circ$, see Fig. ~\ref{fig:3}. By construction, there is one site in the middle of each layer forming a single cluster with coupling $t_\perp$. In contrast to the commensurate case, in the incommensurate case, due to the aperiodicity of the lattice, the other two-site clusters 
present generally different inter-layer couplings. 
In order to simplify the analysis, we classify the clusters into four families, with respectively $t_{\perp,c=1\dots 4}/t_\perp \in [0.75,1)$, $[0.5,0.75)$, $[0.2, 0.5)$, and $[0.06, 0.2)$, respectively. We then assign to the clusters in each family the average value within the family, i.e. $t_{\perp,c=1\dots 4}/t_\perp = 
0.875$, $0.625$, $0.35$, and $0.013$. Slightly different choices of these cluster families do not change the qualitative picture discussed below.

We consider first the case $V_\perp=0$ and $t_\perp=U$, see Fig. ~\ref{fig:4}. 
The ground-state phase diagram in this case is characterized by a broad Mott insulator 
regime with one particle in all sites, separated from the SF regime by a BG region. 
For larger chemical potentials, we observe the appearance of a wide BG regime at low $t/U$, which 
almost completely wipes out any other Mott-like insulating lobes. That regime is characterized at low $t/U$ 
by the formation of small pockets of sites with $S=1$, formed by sites in clusters $c=4$ and 
unclustered sites, see Fig.~\ref{fig:4}~(c). Note that even at $t=0$, there may be pockets of 
sites in which particles can move. This is because motion between sites may be established solely by inter-layer hops. When $t/U$ increases, inside the BG phase, larger~(but still disconnected) regions with $S=1$ form [see Fig.~\ref{fig:4}~(d)]. As a result in the BG phase, $F>0$, but $P=0$, i.e. there is no overall percolation of the $S=1$ sites, see Fig.~\ref{fig:4}~(b). For an even larger $t/U$ the system eventually enters the SF phase with percolated $S=1$ sites~(a more clear example of this is discussed below). The phase diagram 
resembles qualitatively that obtained for strongly disordered Bose-Hubbard models~\cite{fisher1989boson}.



\begin{figure*}[t!]
\centering
\includegraphics[width=2.0\columnwidth]{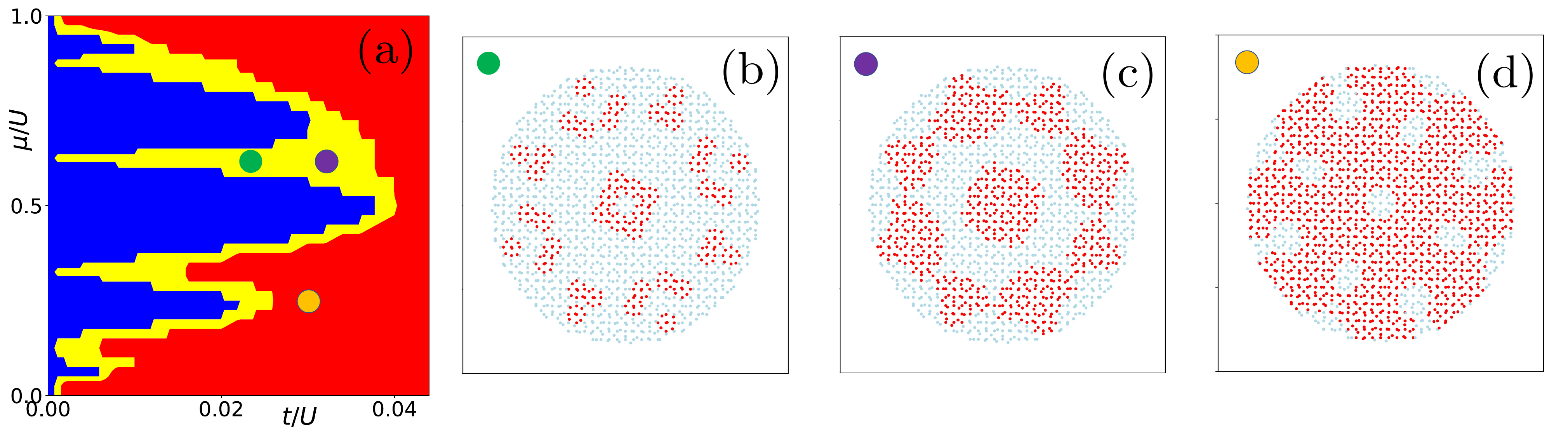}
\caption{
(a) Phase diagram for $\theta = \theta(2,1)+3^\circ$, $V_\perp/U=1$, and $t_\perp=0$. 
Insulating Mott-like phases are indicated in blue, BG phases in yellow, and superfluid phases in red.
(b)--(d) Distribution of the sites with $S=1$~(red) and $S=0$~(blue), for the cases indicated 
in (a) by the colored circles.
}
\label{fig:5}
\end{figure*}

We discuss at this point a different scenario, in which the layers are not coherently coupled, i.e. 
there is no interlayer hopping~($t_\perp=0$). We consider, however, $V_\perp\neq 0$. 
We show in the following that interlayer interactions may result as well in a rich ground-state 
phase diagram. In the absence of inter-layer coupling, we may carry on with the idea of clusterization
but classifying the clusters (still formed by only two sites) according to the inter-layer interaction strength between their sites. 

Note that for commensurate twist angles, unless they are very small, interlayer interactions, which decay much faster than the inter-layer hopping with distance, are typically very weak, except for the case of sites placed on top of each other. As a result, the layers behave as almost disconnected in the absence of interlayer hopping, except for the sites which are on top of each other, in which particles 
may become disconnected from the rest of the lattice if $V_\perp\gg t$. Excluding these sites, however does not result in disconnected regions in each layer, and hence it does not affect the overal mobility in the lattice. As a result BG is not observed. We note that for very small twist angles, interactions may outproject more sites in the lattice at each layer, and as a result a BG may occur, as recently reported in Ref.~\cite{zeng2024dynamical}

In contrast, even for large angles, incommensurate twist angles result in sites which may be eventually close enough to experience 
significant interlayer interactions, and as result, for aperiodic lattices interlayer interactions do significantly affect the phase diagram, even for $t_\perp=0$, resulting 
in a rich ground-state physics. We illustrate this physics again for the case of $\theta(2,1)+3^\circ$, 
with $t_\perp=0$ and $V_\perp/U=1$, see Fig.~\ref{fig:5}~(a).

As for the case of interlayer hopping, pairs of neighboring sites in each layer present 
very different interlayer interactions across the lattice. We hence split the sites into 
different cluster families, according to their interlayer interactions, into 5 families with 
$V_{\perp,c=1\dots 5}/V_\perp \in 1$, $[0.75,1)$, $[0.5,0.75)$, $[0.2, 0.5)$, and $[0.06, 0.2)$. As for the case with $t_\perp>0$
we characterized the phases with the parameters $F$ and $P$, discussed above. 
For the case considered we observe several Mott-like insulating phases, which we characterized by the 
number of particles per site in the different clusters. The three largest Mott-like regions 
are characterized, from bottom to top, by populations $(1,1,1,2,2)$, $(1,1,2,2,2)$, and $(1,2,2,2,2)$. Note 
that the clusters with lower interlayer interactions are more populated, as expected 
due to the repulsive character of the interactions. The BG regime completely surrounds the Mott-like 
regions, separating them from the SF regime. Note that mobility results exclusively from intra-layer hops, and hence in contrast to the case of finite $t_\perp$, 
at $t=0$ no mobility is possible and the phase diagram is dominated by the Mott-like regions.
Figs~\ref{fig:5}~(b) and~(c) show the spatial distribution of $S=1$ sites within the BG regime. 
For growing $t/U$ the islands of sites in which particles can move become larger, although in the BG regime they remain disconnected. In the SF regime, the regions with $S=1$ percolate, but islands of 
insulating sites (which do not participate in particle transport) remain in the system, see Fig.~\ref{fig:5}~(d). The latter case is characterized by $P=1$~(i.e. percolation), but $F<1$, i.e. not all sites have $S=1$.


\section{Conclusions}
\label{sec:Conclusions}

 Cold atoms in twisted-bilayer optical potentials constitute an excellent platform for the study of the interplay of interactions and the twisted geometry, since they present vastly tunable parameters, a feature typically not shared by the solid-state implementation of similar setups. Our results illustrate the rich and highly non-trivial ground-state physics introduced by interactions in twisted-bilayer-like optical lattices, which results from the interplay between inter- and intra-layer hopping and interactions. Crucially, inter-layer hopping induces the formation of two-site clusters, which we have properly taken into account by means of a specifically tailored cluster Gutzwiller approach.
For commensurate twist angles~(periodic lattices) regular cluster families with different interlayer couplings are formed. Cluster formation results in a rich variety of Mott-like phases. Moreover, even for 
periodic lattices, Bose-glass-like phases occur in which particles can move in between 
specific sites forming isolated pockets, without establishing an overall superfluidity. 
The BG-like regime is significantly more relevant in the case of incommensurate twistings~(aperiodic lattice). 
Interestingly, for that case (but not for the commensurate one, at least for large angles), highly non-trivial BG-like and superfluid 
phases may result exclusively from inter-layer interactions, even in the absence of interlayer hopping. 
Although we have focused on ground-state properties, a highly non-trivial dynamics is also expected in these setups, which may be studied as well using the introduced cluster Gutzwiller formalism.


 \acknowledgments{GCP thanks Dean Johnstone for discussion regarding percolation analysis. We acknowledge support of the Deutsche Forschungsgemeinschaft (DFG, German Research Foundation) under Germany's Excellence Strategy -- EXC-2123 Quantum-Frontiers -- 390837967.}

\appendix

\bibliography{Interacting-TBG}

\end{document}